\documentclass[prb,reprint,aps,floatfix,showpacs,superscriptaddress,longbibliography]{revtex4-1}
\usepackage{amssymb}
\usepackage{bbm}
\usepackage{dsfont}

\usepackage{amsmath}
\usepackage{graphicx}
\usepackage[caption=false]{subfig}
\usepackage[colorlinks=true,linkcolor=blue,anchorcolor=red,citecolor=blue,urlcolor=blue]{hyperref}
\begin{document}
\title{Persistent Friedel oscillations in Graphene due to a weak magnetic field}
\author{Ke Wang}
\email{kewang@umass.edu}
\affiliation{%
	Department of Physics, University of Massachusetts, Amherst, MA 01003, USA}
\author{M. E. Raikh}
\affiliation{Department of Physics and Astronomy, University of Utah, Salt Lake City, UT 84112, USA}
\author{T. A. Sedrakyan}%
\email{tsedrakyan@umass.edu}
\affiliation{%
	Department of Physics, University of Massachusetts, Amherst, MA 01003, USA}

\date{\today}
\begin{abstract}

Two opposite chiralities of Dirac electrons in a 2D graphene sheet modify the Friedel oscillations strongly: electrostatic potential around an impurity in graphene decays much faster
than in 2D electron gas. At distances $r$ much larger than the de Broglie wavelength, it decays as $1/r^3$. Here we show that a weak uniform magnetic field affects the Friedel oscillations in an anomalous way.
It creates a field-dependent contribution which is {\em dominant}
in a parametrically large spatial interval $p_0^{-1}\lesssim r\lesssim k_Fl^2$, where $l$ is the magnetic length, $k_F$ is Fermi momentum and $p_0^{-1}=(k_Fl)^{4/3}/k_F$. Moreover, in this interval, the field-dependent oscillations do not decay with distance. The effect originates from a spin-dependent magnetic phase accumulated by the electron propagator.  The obtained phase may give rise to novel interaction effects in transport and thermodynamic characteristics of graphene and graphene-based heterostructures. 
 

%
\end{abstract}
\maketitle

%
\section{introduction}
Graphene, a single-atom-thick honeycomb sheet of carbon atoms\cite{sc04Novoselov,nature05Novoselov}, possesses unusual material characteristics as compared to other 2D electron systems. Having high electron mobilities, the particles in Graphene are orders of magnitude faster than in silicon; they conduct heat much more efficiently than in diamond and conduct current order of magnitude better than in copper. Among other unique properties, Graphene is transparent and impermeable to most gases and liquids, including helium\cite{nature20Sun}. It is harder than diamond and more elastic than fiber carbon at the same time.

Unique electronic properties of Graphene\cite{nature05Kim,prl05Gusynin,iop06Wunsch,prl06Altland,prl06Cheianov,prl06Aleiner,beenakker3,beenakker1,beenakker2,prl07Kim,prb07Mariani,prl08Tse,prb08Bena,prb09Bena,rmp09Castroneto,nature09Kim,prb10Bena,pr10Vozmediano,prb10Virosztek,prl11Stauber,rmp11goerbig,prb11Vaishnav,rmp11Dassarma,rmp12Kotov,IOP12Gordon,np12Nandkishore,prb13Bena,prb13Lawlor,pr16Amorim,prb18Rusin,nature19Dutreix,arxiv20Agarwal,arxiv20Sedrakyan,prb19maiti,nature20Sun} stem from the fact that it is single-atom-thick.
It supports carriers with Dirac-like dispersion.
When doping is low, the Fermi level is located in the close vicinity of $K$ and $K'$ points in the Brillouin zone. The reason for this is that Graphene quasiparticles possess chiral properties related to the two-sublattice structure of the honeycomb lattice. The latter also implies that the lattice unit cell contains two sites (atoms), leading to a ``pseudospin" degree of freedom.


 One of the most prominent effects in regular 2D electron systems
is the interaction-induced zero-bias anomaly in the tunnel density of states (DOS). For small
impurity concentration, this anomaly can be traced to the fact\cite{prb97Glazman}
that impurities are dressed with
Friedel oscillations of the electron density\cite{52Friedel} which falls off as $1/r^2$ with distance, $r$,
from the impurity. Modification of the wave functions due to scattering of electrons from the
dressed impurities gives rise to the singular correction to the self-energy. Upon the advent
of Graphene, the calculations similar to that in 2D gas,\cite{prl06Cheianov,iop06Wunsch,prb07Mariani} indicated
that the zero-bias anomaly in Graphene is absent. The
underlying reason for this absence is that, with matrix underlying Hamiltonian of spin-orbit type,  the backscattering of electrons is forbidden\cite{prb99Raikh}. As a result, the Friedel oscillations in Graphene falls off
faster than in 2D gas, as $1/r^3$.

Since Refs. \onlinecite{prl06Cheianov,prb07Mariani,iop06Wunsch} the Friedel oscillations in Graphene were studied
in tiniest details, both analytically within continuum approximation and numerically, within tight-binding approximation. The results are summarized in the review\cite{physique16Bena}.

Application of a magnetic field turns the spectrum of Graphene into a ladder of non-equidistant Landau
levels. The corresponding perturbation of the electron density around an impurity can be cast into a
sum over these levels\cite{physique16Bena,prb18Rusin}. Still, at elevated temperatures, the
discreteness of the Landau levels does not manifest itself, and the behavior of the Friedel
oscillations with distance becomes quite a nontrivial issue.
A natural expectation is that a weak, non-quantizing magnetic field, modifies the Friedel
oscillations in Graphene in the same way as in the 2D electron gas\cite{prl07Sedrakyan}.
By causing the curving of the semiclassical trajectories, the field gives rise to the
position-dependent magnetic phase, and, thus, breaks the periodicity of the oscillations.
Still, the decay law of the oscillations remains the same as in a zero field.
In fact, such an intuitive reasoning, in application to Graphene, is wrong. It is not only the phase but also the {\em magnitude} of the Friedel oscillations that exhibits a crucial dependence on the magnetic field.

In the present paper we consider this question systematically
and find the field-dependent form of the Friedel oscillations.
We shed light on the nature of the magnetic field modification.
Our key finding is that the potential oscillates anomalously.
Namely, it
{\em does not fall off} with distance, $r$, in a
parametrically large interval.
This omnipresent effect plays a central role in a variety
of quantum many-body contexts in Graphene.  The polarization operator (PO) is an essential quantity for the evaluation of interaction effects using the Feynman diagrams. The non-decaying part in the PO dramatically changes the power counting in the integrand expressing Feynman diagrams. From the new power counting, the magnetic effect may give rise to a new zero-bias anomaly in DOS of Graphene, modify to the quasi-particle lifetime and thermodynamics of Dirac electrons in the Fermi liquid regime. It may also induce new temperature dependence for the dc/ac conductivities\cite{unpublished}.   
The obtained non-decaying Friedel oscillations open an avenue for controlled studies of magnetotransport. They also manifest themselves in field-related thermodynamic properties of Graphene and materials with a pseudo magnetic field such as randomly strained Graphene, stacked and twisted Dirac materials, and the properties of the wormholes in them\cite{npb10Herrero,sym20Capozziello,npb20Garcia}.

The Hamiltonian that incorporates the $B$ field in Landau gauge reads $H=H_{B}+\hat{u} V_{\text{imp}}(r)$,
\begin{eqnarray}
\label{Hamiltonian}
\hat{H}_{B}= v_F \left[  (p_x-eBy) \hat{\Sigma}_x+p_y \hat{\Sigma}_y \right],
\end{eqnarray}
where $V_{\text{imp}}(r)$ is the short-ranged impurity potential,
 $v_F$ is the Fermi velocity and $\hat{\Sigma}_{x,y}=\hat{\sigma}_{x,y} \otimes \hat{\tau}_z$. One can define $\hat{\Sigma}_{z}=\hat{\sigma}_{z} \otimes \hat{\tau}_0$, together with $\hat{\Sigma}_{x,y}$, to form a su(2)-algebra. The Pauli matrices $\hat\sigma_{x,y,z}$ act in the space of $A$ and $B$ sublattices of the honeycomb lattice, $\hat{\tau}_z$ is the Pauli matrix distinguishing between two Dirac points ( $K$ and $K'$ ) of the Graphene dispersion relation and $\hat{\tau}_0$ is the identity matrix.  We consider the simplest case of
the diagonal disorder $\hat{u}=u\hat{I}$, where $u$ is a scalar. 
The uniform field breaks the chiral symmetry near each Dirac point individually since, $\mathbf{p}\cdot \mathbf{\hat\Sigma}$, $\hat\Sigma=(\hat\Sigma_x, \hat\Sigma_y),$
does not commute with the Hamiltonian. 
It is this non-commutativity,
specific for graphene and other Dirac materials\cite{aip14Balatsky}, that causes the
observable modification of Friedel oscillations in a weak magnetic field, $B$. Quantitatively,
the criterion of weak field is that the
the magnetic length, $l=(\frac{\hbar c}{eB})^{1/2}$, is much larger than the de Broglie wavelength, $k_F^{-1}$. Below we show that weak field modifies the screened  Coulomb potential\cite{prb93Yue,prb01Zala,prb05Adamov} $V(r)$ to
\begin{figure}
	\includegraphics[scale=0.5]{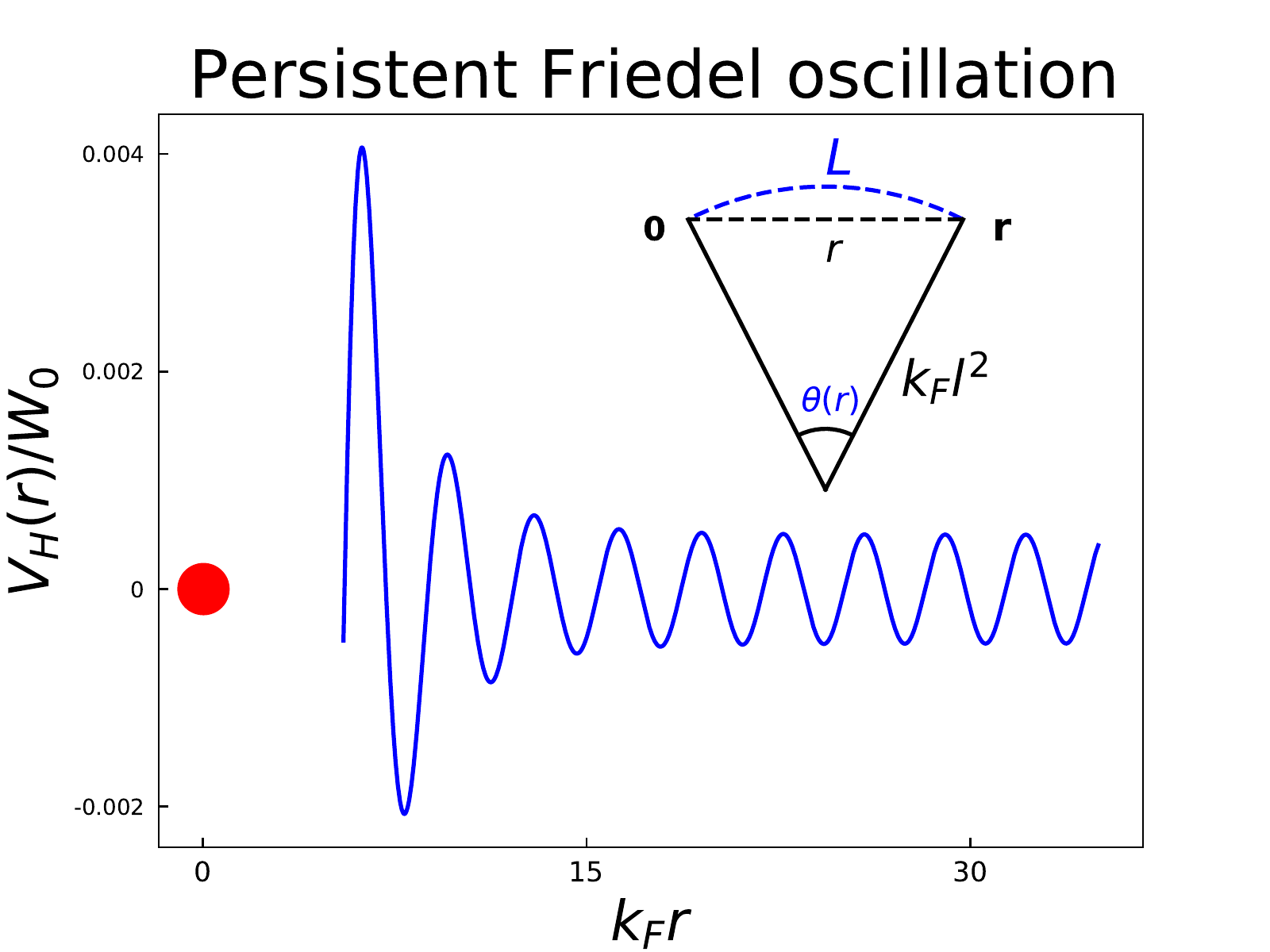}
	\centering
	\caption{(Color online)  Electrostatic potential, $V_H(r)$, is plotted vs. dimensionless distance $k_Fr$ from the impurity  [big (red) circle] in the presence of a weak magnetic field, $B$, in the range $1\ll k_F r\ll k^2_F l^2$. The figure is obtained from Eq.~(\ref{2}) using a typical value of $p_0/k_F=0.1$. The potential $V_H(r)$ is measured in units of $W_0= {k^3_F g V(2k_F)}/{2\pi^2 v_F}$. The amplitude of oscillations first decays as  $1/r^3$ and then converges to a constant $\propto B^2$.    The inset illustrates the classical trajectory of 2D electrons between $0$ and $ \mathbf{r}$ in the presence of a weak magnetic field. $L$ is length of the arc, $r$ is the simply $|\mathbf{r}|$ and $\theta(r)$, the angle of the arc, is approximately given by $r/k_F l^2$. } 
	
	\label{FO}
\end{figure}
\begin{eqnarray}
\label{2}
V_H(r)=\frac{g{V}( 2k_F) }{2\pi^2v_F r^2}
\Biggl[\frac{1}{r}\cos\Bigl(2k_Fr-\frac{p^3_0 r^3}{12}\Bigr)
\nonumber\\
+\frac{r^2}{2k_F l^4}\sin\Bigl(2k_Fr-\frac{p^3_0 r^3}{12}\Bigr)
\Biggr].
\end{eqnarray}
 Here $V(2k_F)$ is the $2k_F$ component of the interaction, while the impurity potential is treated in the Born approximation with $g=u\int d^2r V_{\text{imp}}(r)$. There are two competing terms. One can see that when the magnetic phase $p^3_0 r^3 \ll 1$ with $p_0^{-1}=(k_Fl)^{4/3}/k_F$, the potential is decaying as a polynomial function, $\sim1/r^3$. When $1 \ll p^3_0 r^3 \ll k_F r$, the potential is oscillating anomalously, with a {\em constant} amplitude. This persistent effect comes from the diagonal part $\hat{u}=u\hat{I}$ of impurity potential, while other non-magnetic impurity potentials do not contribute to $V_H(r)$ in the leading order in impurity scattering. (for details, see Appendix~\ref{rs} and ~\ref{disorder}.)  
 
The paper is organized as follows. In Sec.(\ref{sp}), we present a qualitative derivation of persistent Friedel oscillations, which follows from the semi-classical magnetic phase accumulated by the electron propagator. In Sec.(\ref{PO}), we present a thorough calculation of the polarization operator in the presence of a weak magnetic field and derive Friedel oscillations of electron density. The implications to interaction effects are discussed in Sec.(\ref{in}). Concluding remarks are given in Sec.(\ref{cm}).

\section{ Qualitative discussion} \label{sp}
In this section, we make a qualitative argument for new effects of Friedel oscillations in Eq.~(\ref{FO}). Both modifications of Friedel oscillations, namely the phase $p_0^3 r^3$ in oscillations and the persistent part, can be understood semiclassically. 
The electron propagator, $G_{s,s'}(0,\mathbf{r})$, where $s=\pm$  represent $A/B$ sublattices, accumulates a phase, $k_{s,s'}L$, when electrons propagate along the semi-classical arc. Here $L$ is the length of arc shown in the inset of Fig.~(\ref{FO}) and $k_{s,s'}$ is an effective momentum.
 The diagonal component of Dirac propagator, $G_{s,s}(0,\mathbf{r})$, can be understood as  a propagator of the 2D electron gas with an effective Fermi energy $E^s_F=E_F[1-s(2k^2_Fl^2)^{-1}]$ and an effective cyclotron frequency $\omega_0=v_F (k_Fl^2)^{-1}$.
This yields an effective momentum $k_{s,s}=k_F[1-s(2k^2_Fl^2)^{-1}]$ for diagonal propagators. While if $s\neq s'$, $k_{s,s'}=k_F$. For details of deriving effective momentums, see Appendix~(\ref{effective}).

The phase $p_0^3 r^3$ in the oscillations is due to the curving of the path. Semiclassically, the propagator acquires a magnetic phase $k_F (L-r)$ because of the curving of trajectory. Since $L= k_Fl^2 \theta$, $r=2k_Fl^2\sin(\theta/2)$ and $\theta(r)\simeq r/k_F l^2$, the magnetic phase $k_F (L-r)$ becomes equal to $p_0^3 r^3/24$. The PO involves a product of two propagators, and thus the magnetic phase in PO is doubled.  This is exactly the magnetic phase of Friedel oscillations. 

The persistent part of Friedel oscillations originates from the deviation of $k_{s,s'}$ from $k_F$. The effective momentum implies spin-dependent magnetic phases, $\sim(k_{s,s}-k_F)r=-sr (2k_F l^2)^{-1}$  of diagonal propagators . Although the phase is spin-dependent, it does not depend on the valleys. The phase could then be expressed compactly, $- \theta(r)\hat\Sigma_z /2$. The PO involves a trace of two propagators. Using the fact that Pauli matrices are traceless  and $\textbf{tr}\hat\Sigma_z^2\neq 0$, the leading magnetic contribution to PO is from a square of $\theta(r)$, namely $\theta(r)^2 \propto r^2/k_F^2 l^4$. This is the new amplitude of the second term in Eq.~(\ref{2}).

 Importantly, the anomalous effect in Eq.~(\ref{2}) persists even at high temperatures, $T\sim T_0 \equiv v_Fp_0$, which is much higher than the cyclotron energy. The temperature scale, $T_0$ can be derived qualitatively. Quasi-classically, the electron propagator accumulates a dynamical phase $k_F L$ along the arc (see the inset in Fig.~\ref{FO}). The condition $k_F (L-r)\sim 1$ can be cast in the form  $r\sim p_0^{-1}$, since $k_F (L-r)\propto (p_0 r)^3$. Then, the spatial scale $p_0^{-1}$ translates into the energy scale $v_F p_0$. This, in turn, sets the temperature scale, $T_0$. 





\section{The Polarization operator} \label{PO}

In this section, we show the emergence of persistent Friedel oscillations by calculating the PO rigorously in the momentum space. We start from the summation over Landau levels for PO and then develop a low energy effective theory around the Fermi level. We show how the spin-dependent magnetic phase of the gauge-invariant electron propagator manifests itself in the matrix elements of PO. Using the obtained form of the PO, we derive the Friedel oscillations and observe their persistent behavior. Finally, we discuss the smearing of anomalies in the PO under a weak magnetic field.

 \subsection{Summation over Landau levels}.
We start from a general expression for the PO in the momentum space
\begin{equation}
\label{Pi}
\Pi(k,\omega)=\sum^\infty_{n,n'=0}\sum_{s,s'=\pm}
\frac{n_F(s'\omega_{n'})-n_F(s\omega_n)}{\omega-(s\omega_n-s'\omega_{n'})}\Big\vert M_{s,s',k}^{n,n'}\Big\vert^2,
\end{equation}
where $n_F(\omega)$ is the Fermi-Dirac distribution,
while the frequencies, $\omega_n$, are given by
$\omega_n=\left(2n\right)^{1/2}v_F/l$. The quantities $M_{s,s',k}^{n,n'}$
are the matrix elements of $\exp\left(i{\bf k r}\right)$ between the
states $\langle s,n\vert$ and $\vert s',n'\rangle$. Since the wave functions
are the vectors consisting
of the oscillator states $n$ and $n+1$, the square of the matrix element
can be expressed via
the generalized Laguerre polynomials, $L^n_m$, as follows
\begin{eqnarray}
\label{M}
\Big\vert M_{s,s',k}^{n,n'}\Big\vert^2=(-1)^{n'-n}\frac{e^{-x} }{   \pi l^2}
\Bigl[L^{n-n'}_{n'-1} (x) L^{n'-n}_{n-1}(x)\nonumber\\
+  L^{n-n'}_{n'} (x) L^{n'-n}_{n}(x)  +2ss'\left(\frac{n}{n'}\right)^{1/2} L^{n-n'}_{n'-1} (x) L^{n'-n}_{n}(x)  \Bigr].
\nonumber\\
\end{eqnarray}
where $x=k^2l^2/2$.
The summation in Eq. (\ref{Pi}) is performed over two valleys and two spins.
However, the main contribution comes from the states near the Fermi level, $E_F$, which we
assume to be positive. This allows to set $s=s'=1$ in Eq. (\ref{M}).
The condition that the magnetic field is weak can be cast in the form, $N_F\gg 1$, where
$N_F= k^2_Fl^2/2   $ is the number of Landau levels with energies between $\epsilon=0$
and $\epsilon=E_F$.

To perform the summation over $n$ and $n'$ it is convenient to use the following
integral representation of the  Laguerre polynomials
\begin{equation}
\label{representation}
\!L_m^n(x)\!=\!\frac{1}{2\pi} \int\limits_0^{2\pi} \frac{d\theta}{(1-e^{i\theta})^{n+1}} \exp \left\{\frac{xe^{i\theta}}{e^{i\theta}-1}-im\theta\right\}.
\end{equation}
In the vicinity of the  Kohn anomaly, $k\approx 2k_F$, we have $x\gg 1$.
Under this condition, the major contribution to the integral
Eq. (\ref{representation}) comes from the vicinity of $\theta =\pi$.
Substituting $\theta=\pi+\psi$ into the integrand and expanding with respect
to $\psi$ yields
the following integral representation for
the square of the matrix element (details see Appendix~\ref{matrix})
\begin{widetext}
\begin{eqnarray}
\label{simp_M}
&&\Big\vert M_{s,s,k}^{n,n'}\Big\vert^2=\frac{1}{4\pi^3 l^2}\int \frac{d\psi d\psi'}{4}
\exp{\Big\{\frac{ix}{48}\left(\psi^3+\psi'^3 \right)\Big\}}
\nonumber  \Biggl[ \sum_{\nu=\pm}
\exp    \Big\{i\left( \frac{x}{4}- \frac{n+n'+\nu}{2} \right)(\psi+\psi') \Big\}\\
&-2&\exp    \Big\{i\left( \frac{x}{4}- \frac{n+n'}{2}\right)(\psi+\psi') +i\frac{\psi-\psi'}{2} \Big\} \Biggr].
\end{eqnarray}
\end{widetext}
Here the spin-dependent magnetic phase, the signal of chiral symmetry breaking,  manifests itself in the matrix element as a small, but non-negligible, phases $\nu(\psi+\psi')$, with $\nu=\pm $. The negative sign in the second line is the result of the Berry phase $\pi$, which is specific for Dirac electrons. The two features are responsible for the main result of the paper. 

Since the main contribution to the sum in Eq. (\ref{Pi}) comes from $n$ and $n'$ close to $N_F$,
it is convenient to introduce the new variables
$m=N_F-n$, $m'=-N_F+n'$. Then the summation in Eq. (\ref{Pi})
can be performed with the help of the following identity
\begin{eqnarray}
\label{identity}
&& \sum_{m,m'=-\infty}^{+\infty} \frac{ n_F(\epsilon_F+\frac{\sqrt{2}v_F}{2l}m' )-n_F( \epsilon_F-\frac{\sqrt{2}v_F}{2l}m)}{m'+m } \nonumber\\
&& \times    \cos \left[     (m'-m)  \alpha+\beta               \right]
=-\frac{ \pi^2 T \cos \beta }{\omega_0 \sinh (2\pi |\alpha|T/\omega_0)},
\end{eqnarray}
where $\omega_0= v_F (k_F l^2)^{-1}$ is the effective cyclotron frequency and $\alpha$, $\beta$ are real numbers.   When applying the above identity to  the summation in Eq.~(\ref{Pi}), we set
	$\alpha=y$ with $y\equiv 2^{-1}\left(\psi+\psi'\right)$ and $\beta=0$. As we will see in the next section, the integration over $y$ defines a characteristic scale, $y\sim   (k_F l)^{-2/3} $. This scale for $y$ implies that the temperature damping term $A(T)\equiv T \left[\omega_0 \sinh (2\pi y T/\omega_0)\right]^{-1}$ is essentially temperature independent  at $T\ll T_0$, namely $ A(T)\approx  \left[2\pi (k_F l)^{2/3}\right]^{-1}$.  At $T\gg T_0$, the damping factor is important, as it becomes exponential: $ A(T) \approx 2T \omega^{-1}_0 \exp{\left(- \pi  T/T_0\right)}$. In the following, we work in the low temperature limit, $T\ll T_0$. This effect of persistent oscillation will survive up to $T\sim T_0$, while at higher temperatures  the Friedel oscillations will be washed out.

\subsection{  The form of the PO} Equipped with Eq.~(\ref{simp_M}) and ~(\ref{identity}), one can write the static PO, $\Pi( k)\equiv \Pi(k,0)$, as a single integral with respect to variable $y $. Details see Appendix~(\ref{integral}). To clarify two different effects in $\Pi( k)$, we present $\Pi( k)$ as sum of two terms $\Pi_{1}( k)+\Pi_{2}( k)$. Here $\Pi_{1}(k)$ and $\Pi_{2}( k)$ are expressed by
\begin{eqnarray}
\label{Pi_one}
&&\Pi_{1}(k)=-\frac{1  }{ 4 \pi^{3/2}v_F l } \int\limits_a^{\infty}\frac{dy}{y^{3/2}}  \\
&\times&\Biggl\{ \sum_{\nu=\pm 1}\cos \Biggl[  \left( k_F\delta k l^2-\nu\right)  y+\frac{k^2_Fl^2y^3}{12} +\frac{\pi}{4}\Biggr]  \nonumber\\
&-& 2\cos \Biggl[    k_F \delta k l^2 y+\frac{k^2_Fl^2y^3}{12} +\frac{\pi}{4} \nonumber\Biggr] \Biggr\},
\end{eqnarray}
and
\begin{eqnarray}
\label{Pi_second}
&&\Pi_{2}(k)=\frac{1  }{ 2 v_F l\pi^{3/2} } \int\limits_a^{\infty}\frac{dy}{y^{3/2}} \Biggl(\frac{1}{k^2_Fl^2y}\Biggr)\nonumber \\
&\times&   \sin \Bigl(    k_F\delta k l^2   y+\frac{k^2_Fl^2y^3}{12}+\frac{\pi}{4}\Bigr)  .
\end{eqnarray}
Here $\delta k=k-2k_F$ is the momentum measured from $2k_F$. Finite low-$y$ cutoff, $a$, of the order of the lattice spacing,
does not affect the form of the Friedel oscillations. We will see that $\Pi_1$ and $\Pi_2$ are responsible
for the two distinct contributions to the Friedel oscillations
in Eq.~(\ref{2}).


{Here we derive the PO in the real space.} We start with the contribution  Eq. (\ref{Pi_one}).
Transformation  to the real space
is accomplished by the following radial integral 
\begin{eqnarray} 
\Pi(r)= \int_0^{\infty} kdk (2\pi )^{-1} J_0 (k r) \Pi(k)\nonumber \\ \simeq   k_F\int_{-\infty}^{\infty} d\delta k (2\pi )^{-1} J_0 (k_F r) \Pi(k_F+\delta k)
\end{eqnarray}
where $J_0(x)$ is the zeroth-order Bessel function. Here we have used the fact that $\delta k \ll k_F$. In the domain, $k_Fr\gg 1$, we can replace the Bessel function by the large-$x$ asymptote $J_0(x)\approx \left(2/\pi x \right)^{1/2}\cos \left(x-\frac{\pi}{4}\right)$. The integration over $k$ sets $y=r (k_Fl^2)^{-1}$. Then the summation over $\nu$ in Eq.~(\ref{Pi_one}) yields
\begin{eqnarray}
\label{simplified}
\Pi_{1}(r)&=&-   \frac{k_F  }{ 2 v_F  \pi^{2} r^2 } \sin \Bigl(    2k_Fr  -\frac{p_0^3r^3}{12}\Bigr)\nonumber\\
&\times& \Biggl[\cos \Bigl( \frac{r}{k_Fl^2}\Bigr)  -1\Biggr].
\end{eqnarray}
 The effect of the weak magnetic field is not negligible if the magnetic phase $p_0^3r^3\sim 1$. From here, the characteristic scale for $y$ is $ (p_0k_Fl^2)^{-1}= (k_F l)^{-2/3}$.  
Since we consider the distances  $ r \ll k_F l^2$, i.e.   much smaller than the
Lamour radius, the magnitude of $2k_Fr$ oscillations given by $\sin$-function in the equation above can be further simplified to  $k_F (2 v_F  \pi^{2} r^2 )^{-1} (1-\cos r/k_Fl^2 )= (4 v_F  \pi^{2} k_F l^4 )^{-1} $.
The result for $\Pi_1$ describes the contribution to the oscillations of the electron density
which do not decay with distance in the domain $k_F^{-1}\ll r \ll k_Fl^2$. It reproduces the second term in Eq. (\ref{2}).

 Evaluation of the contribution $\Pi_{2}(r)$
to the PO defined by Eq. (\ref{Pi_second}) involves the same
steps as evaluation of $\Pi_{1}(r)$.  The result reads
\begin{eqnarray}
\label{14}
\Pi_{2}(r)&\approx&\frac{1 }{2\pi^2v_Fr^3} {\cos\Bigl(2k_Fr-\frac{p^3_0 r^3}{12}\Bigr)}.
\end{eqnarray}
It reproduces the first term in Eq. (\ref{2}). The decay $1/r^3$ is specific for Graphene,
while the phase is the same as in 2D electron gas. 

The the real space static PO, $\Pi(r)$, determines the Hartree potential $V_H(r)$, via modulation of the electron density $\delta n(\mathbf{r})$ around the impurity. Within the Born approximation, $\delta n(\mathbf{r})=g\Pi(r)$. Since the density modulation originates from the backscattering of fermions,  $\delta n(\mathbf{r})$ determines the Hartree potential as $V_H(r)= V(2k_F) \delta n(\mathbf{r})$ (The derivation see Appendix~\ref{Hartree}). As such, the Hartree potential $V_H(r)$ is equivalent to $gV(2k_F)\Pi(r)$.

 \begin{figure}
	\centering
	\includegraphics[scale=0.5]{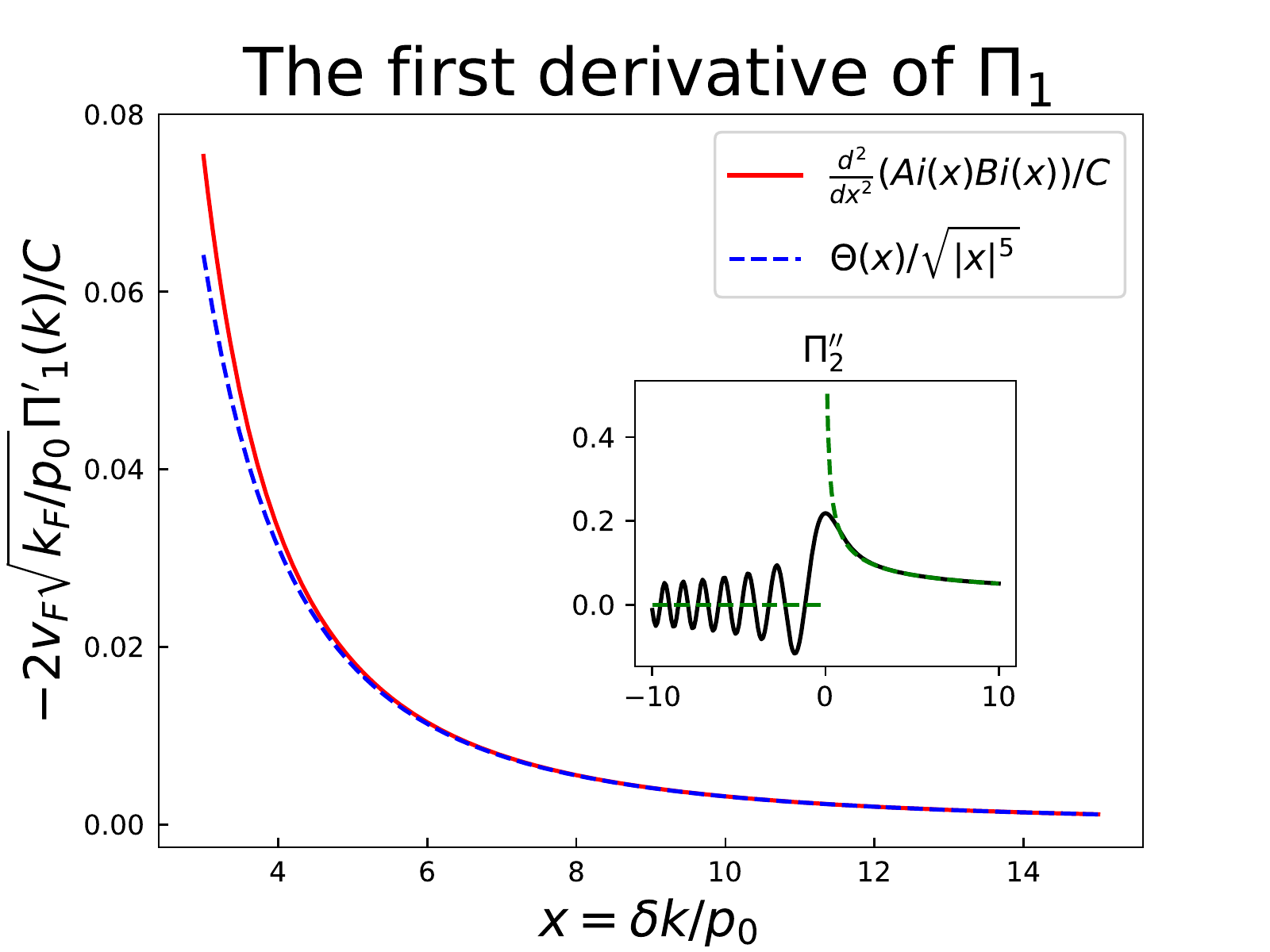} 
	\caption{(Color online) 
Asymptote of derivative $\Pi'_1(k)$ is plotted versus the dimensionless $x=\delta k/p_0$. The red curve depicts $-{2v_F}{\sqrt{k_F/p_0}} \Pi'_{\text{1}}(k)/C=[Ai(x)Bi(x)]''/C$ versus $x$. Here $C$ is a positive constant, approximately equal to $0.12$. The blue curve is
		$1/(\sqrt{|x|})^5$, converging to $\Pi'_1$ when $x\gg 1$. The inset depicts $\Pi_2''(k)$. The black curve is $-{v_F}\sqrt{p_0k_F} \Pi''_{2}(k)=Ai(x)Bi(x)$, demostrating the smearing of the anomaly.
		The green represents $D\Theta(x)/\sqrt{|x|}$, namely the Kohn anomaly in PO in a zero field. D is a positive constant, approximately equal to $0.16$.}
	\label{plotpi1} 
\end{figure}

 
\subsection{  Smeared anomalies} \label{sa}
The spin-dependent magnetic phase, $\theta(r)\hat\Sigma_z /2$, leads to a new term, $\Pi_1(k)$, in the PO, while the curving of path smears the existing anomaly in $\Pi_2(k)$. We start from the momentum-space representation of the PO given in terms of the product of the Airy functions\cite{1953Bateman}.
Then we differentiate Eq. (\ref{Pi_second}) with respect to $\delta k$ twice and obtain 
\begin{eqnarray}  \Pi''_{2}(k)=- ( v_F \sqrt{p_0 k_F})^{-1}F ( {\delta k}/{p_0} ) \end{eqnarray}  where $F(z)\equiv\text{Ai}(z)\text{Bi}(z)$. This represents the smearing of the Kohn anomaly of the PO by the weak field. Inset of Fig.~(\ref{plotpi1}) illustrates how the anomaly get smeared.

Importantly, $\Pi_1'(k)$ can also be obtained
as (details see Appendix~\ref{momentum})
\begin{eqnarray} \Pi_1'( k)= (2 v_F )^{-1}\sqrt{p_0/k_F} F''\Bigl( {\delta k}/{p_0}\Bigr). \end{eqnarray}  This term only emerges in the presence of the magnetic field. In the limit $\delta k \gg p_0$, $\Pi_1'(k)$, it converges to zero as 
$\propto B^2 (\delta k)^{-5/2}$. This asymptote, plotted in Fig.~(\ref{plotpi1}), has same origin with persistent oscillations. 	To better understand the effect, one can Fourier transform $\Pi_{\text{1}}(k)$ using the asymptote. From power counting, it is straightforward to see the emergence of the non-decaying oscillating function, $\sim B^2\sin(2k_Fr)$. 

	
 \begin{figure}
  \centering
  \includegraphics[scale=0.22]{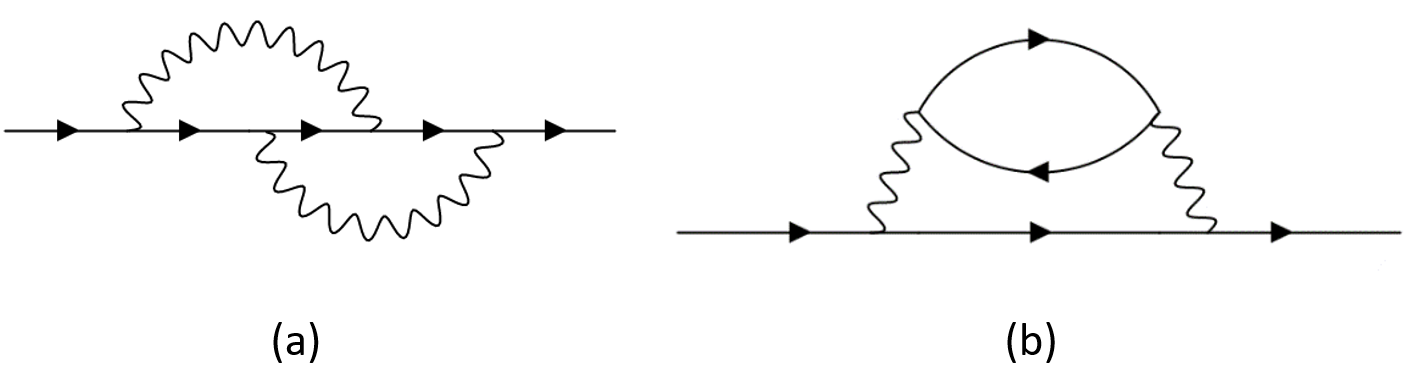} 
  \caption{The leading contributions to the lifetime of quasi-particles. Solid lines represent the Feynman propagators. The wavy lines represent the electron-electron interactions. (a) Fock Digram.  (b) Hatree diagram. }
  \label{diagram}
\end{figure}
\section{  Implications to interaction effects}\label{in}  Graphene is a 2D Fermi-liquid when $E_F>0$. This implies that the scattering rate of the quasi-particle around Fermi surface obeys $\Gamma(\omega) \propto \omega^2/E_F \log(E_F/\omega)$. 
In perturbation theory in interaction parameter, two leading Feynman diagrams contributing to the electron lifetime are shown in Fig.~(\ref{diagram}). We evaluate these diagrams in the presence of the weak magnetic field using the obtained form of the electron propagator. 
The computation leads to an unexpected result\cite{unpublished}: the quasiparticle lifetime acquires a singular in $\omega$ magnetic correction $
\Gamma(\omega;B)-\Gamma(\omega;0)\propto    \omega_0^2/E_F \log\Big(\omega/E_F\Big).
$
Here $\Gamma(\omega;B)$ is the scattering rate of quasi-particle with frequency $\omega$ in the presence of the magnetic field, $B$, $\omega_0=v_F (k_F l^2)^{-1}$. 
The expression is valid when $E_F>\omega>\{\omega_0, T\}$. The magnetic correction above is more singular in frequency $\omega$ than the non-magnetic part. This singularity originates from the spin-dependent magnetic phase, $\theta(r)\hat\Sigma_z$, in electron propagators. Interestingly, the field-dependent interaction corrections to various observables in Graphene,  including zero-bias anomaly in the density of states and ac/dc conductivities, also exhibit more singular behavior in either $\omega$ or temperature, $T$\cite{unpublished}.

\section{  Concluding remarks}\label{cm}  In this paper we demonstrated that weak magnetic field manifests itself in the Friedel oscillations in two ways: it modifies (i) the phase of the oscillations and (ii) makes the magnitude of oscillations
non-decaying in a parametrically large interval.
The origin of the modification of the phase in oscillations, $\sim p^3_0 r^3$,  can be traced to the curving of the classical trajectory of an electron in a weak magnetic field (see inset in Fig.~\ref{FO}).
The trajectory is curved even at $r$ much smaller than the Larmour radius, leading to the magnetic phase\cite{prl07Sedrakyan} $\sim (p_0r)^3$. This effect just by itself leads to remarkable high-temperature interaction effects in 2DEG\cite{prb08Sedrakyan,prl08Sedrakyan,prl08Raikh}.

 Graphene also supports modification of the magnitude of the Friedel oscillations by a weak field. The origin of this effect is an emerging spin-dependent phase in electron propagators, $\sim\exp (-i\hat\Sigma_z \theta(r)/2)$. This effect manifests in persistent Friedel oscillations and leads to non-trivial magnetic corrections to many-body characteristics in Graphene. The transport and thermodynamic properties of monolayer Dirac materials, randomly strained Graphene and stacked and twisted Dirac materials will also be anomalously sensitive to this magnetic phase even at temperatures, $T\sim T_0$, which is much higher than the cyclotron energy\cite{prl07Sedrakyan}. 
 

Technically, to develop the theory of interaction effects in Dirac materials in the presence of the weak field, one can use the obtained form of the PO and/or Friedel oscillations in the Feynman diagrams (in the momentum or real space representations). However, a word of caution is in order here. Since the field dependence also enters the fermion Green's functions, in the Feynman diagrams, one should also consider modified propagators on the same footing along with Friedel oscillations.
The explicit form of the Feynman propagators in the weak B-field is derived in Appendix~\ref{rs}.

Experimentally, Friedel oscillations can be observed with the scanning tunneling microscope (STM), which images 2D surfaces at the atomic level\cite{science97sprunger,nature19Dutreix,prb13Jelena}. In STM, data are determined by backscattering processes along the energy contours.
Experimental tests of these oscillations
would include examining the temperature
dependence of the Friedel oscillations through an
extended range of temperatures $0 \lesssim T \lesssim T_0$, determining the persistent range of oscillations,  $p_0^{-1}\lesssim r\lesssim k_F l^2$, and investigating the effect of their $B$-dependence.

\section{  Acknowledgements}. The research was supported by startup funds from the University of Massachusetts, Amherst (K.W. and T.A.S.), and by the Department of Energy, Office of Basic Energy Sciences, Grant No. DE-FG02-06ER46313 (M.E.R.).




\appendix

\section{Real-space calculation of the Friedel oscillations} \label{rs}
Consider the modification of the Friedel oscillations in graphene by a weak uniform magnetic field. The Hamiltonian that incorporates the $B$ field in Landau gauge reads
\begin{eqnarray}
H_{B}= v_F \left[  (p_x-eBy) \hat{\Sigma}_x+p_y \hat{\Sigma}_y \right]+\hat{u}V_{\text{imp}}(r)
\end{eqnarray}
with $\hat{u}=u\hat{I}$. Here $V_{\text{imp}}(r)$ is the short-ranged impurity potential,
$v_F$ is the Fermi velocity and $\hat{\Sigma}_{x,y}=\hat{\sigma}_{x,y} \otimes \hat{\tau}_z$. One can define $\hat{\Sigma}_{z}=\hat{\sigma}_{z} \otimes \hat{\tau}_0$, together with $\hat{\Sigma}_{x,y}$ to form a su(2)-algebra.
Clearly, the uniform field breaks the chiral symmetry since, in a finite fields, $\mathbf{p}\cdot \mathbf{\Sigma}$ does not commute with  $H_{B}$.
In the following, we report the result for the Green function in the asymptotic region $k^2_F l^2\gg k_F r\gg 1$, $G(\mathbf{x},\mathbf{x}'|\omega)= I(\mathbf{x},\mathbf{x}';\omega)  \hat{M}(\mathbf{r};\omega)$. Here $M$ is given by
\begin{eqnarray}
\label{3}
\hat{M}(\mathbf{r};\omega)\simeq  \Big(\text{sgn}(\omega)+\frac{i}{2k_Fr}  \Big) \hat{r}\cdot \boldsymbol\Sigma\nonumber\\  +\exp\Big\{-i\frac{\theta(r)}{2}  \text{sgn}(\omega)\hat{\Sigma}_z \Big\}
\end{eqnarray}
where  we only preserve the leading orders with the non-quantized condition $k_Fr \ll k^2_F l^2$. Here $l=\sqrt{\hbar/eB}$ is the magnetic length, $k_F$ is the Fermi momentum and $\omega$ is the frequency measured from the Fermi energy. We also define functions $I(\mathbf{x},\mathbf{x'};\omega)$ by
\begin{eqnarray}
&& I(\mathbf{x},\mathbf{x'};\omega)=  \frac{k_F }{2 v_F} \frac{ 1}{  \sqrt{2\pi k_F r  }}  e^{-i\chi}  \\ &&\times  \exp\Big\{i\text{sgn}(\omega)\left({k_Fr}[1+\frac{\omega}{\epsilon_F}]-\frac{p^3_0 r^3}{24}+i\pi/4 \right)\Big\}. \nonumber
\end{eqnarray}
where $r=|\mathbf{x}-\mathbf{x'}|$,  $\chi=(x-x')(y+y')/(2l^2)$ describe breaking of the translational invariance and $p^3_0=1/(k_F l^4)$ is a scale introduced by Sedrakyan, Mishchenko, and Raikh in Ref.~\onlinecite{prl07Sedrakyan} to characterize the  magnetic phase in the Green function of the 2D electron gas.

Having the form of Green's function in the presence of a weak field, one can calculate the polarization operator.
\begin{eqnarray}
\Pi(\mathbf{x},\mathbf{x}';\omega)=&&-i\int_{\Omega} \frac{d\omega'}{2\pi}  {I(\mathbf{x}-\mathbf{x'};\omega')I(\mathbf{x}-\mathbf{x'};\omega'-\omega)} \nonumber\\&& \textbf{tr}  \Big(\hat{M}(\mathbf{r};\omega)\hat{M}(-\mathbf{r};\omega'-\omega)\Big)
\end{eqnarray}
Here we still consider an impurity with $\hat{u}=u\mathbf{I}_4$. 
After putting expression expressions into the equation above, one can obtain the $2k_F$ component of the polarization operator 
\begin{eqnarray}
\Pi_{2k_F}(\mathbf{x},\mathbf{x}';0)&\approx&\frac{u}{v_F}\frac{\cos(2k_Fr-\frac{p^3_0 r^3}{12})}{4\pi^2 r^3} \nonumber\\&&+\frac{u}{v_F}\frac{\sin(2k_Fr-\frac{p^3_0 r^3}{12})}{4\pi^2 } \frac{1}{ 2k_Fl^4} \label{FO1}
\end{eqnarray}
This result fits well with the Eq.~(21) and Eq.~(22) in the main text, the result obtained in the momentum space. The decay of the first oscillatory term here is consistent with the result of free Dirac electrons\cite{iop06Wunsch,prl06Cheianov,prb07Mariani}, however there is an additional magnetic phase $p^3_0 r^3/12$ that breaks periodicity of oscillations. The ratio between magnitudes of the first and the second oscillatory functions is proportional to $k^3_F r^3/k^4_F l^4$. When
(i) $1\ll k_F r \ll (k_F l)^{4/3}$, the polarization operator in real space decays as $1/r^3$; (ii) $ (k_F l)^{4/3} \ll k_F r \ll  (k_F l)^{2}$, the chiral symmetry breaking effect dominates. This symmetry-breaking term is proportional to $B^2 \sin(2k_F r)$, i.e., it oscillates with constant amplitude.

Finally, one can evaluate the effective electrostatic potential around the impurity (where we restore spin 1/2 back), which yields
\begin{eqnarray}
\label{HARTREE}
V_H(r)&=&\frac{g{V}( 2k_F) }{2\pi^2v_F}
\frac{\cos(2k_Fr-\frac{p^3_0 r^3}{12})}{ r^3}\nonumber\\&&+\frac{g{V}( 2k_F) }{2\pi^2v_F} \frac{1}{2}p^3_0{\sin(2k_Fr-\frac{p^3_0 r^3}{12})}
\end{eqnarray}

\section{Effect of disorder} \label{disorder}
In the main text, we considered the disorder potential to be $\hat{u}V_{\text{imp}}$, with $\hat{u}=u\hat{I}$. Around Eq.~(2), we discuss the effects of other disorder potentials on the Hartree potential. This section provides further detailed calculations. Consider a non-magnetic impurity, which should be Hermitian and time-reversal symmetric . Then the $\hat{u}$ could be expressed by ten parameters\cite{prl06McCann,prl06Aleiner}
\begin{eqnarray}
\hat{u}=u\hat{I}+\sum_{s,l=x,y,z} u_{sl } \Sigma_s \Lambda_l
\end{eqnarray}
where the matrices are given by
\begin{eqnarray}
\Sigma_x=\hat{\tau}_z\otimes \hat{\sigma}_x,\quad \Sigma_y=\hat{\tau}_z\otimes \hat{\sigma}_y, \quad \Sigma_z=\hat{\tau}_0\otimes \hat{\sigma}_z  \\
\Lambda_x=\hat{\tau}_x \otimes \hat{\sigma}_z,\quad \Lambda_y=\hat{\tau}_y \otimes\hat{\sigma}_z,\quad \Lambda_z=\hat{\tau}_z \otimes\hat{\sigma}_0.
\end{eqnarray}
Here the Pauli matrices $\hat{\tau}$ are acting on space of the valley indices while the Pauli matrices $\hat{\sigma}$ act in the space of A/B sublattices. Components of impurity potentials are explained as follows:
\begin{itemize}
	\item $u\hat{I}$ is the diagonal disorder, the electric-static potential averaged over A/B sublattices.
	\item $u_{xz}$ and $u_{yz}$ introduce the hoppings between A and B.
	\item $u_{sx}$,$u_{sy}$ introduce the intervalley scatterings, $s=x,y,z$.
	\item $u_{zz}$ defines the difference between on-site chemical potentials on A and B sublattices. It represents the sublattice symmetry breaking mass term.
\end{itemize}
The polarization operator is given by
\begin{eqnarray}
\Pi(\mathbf{r},\mathbf{r}';\omega)&=&-i\text{tr}\Big(\hat{u} \int_{-\infty}^{+\infty} \frac{dw'}{2\pi}   G(\mathbf{r},\mathbf{r}';\omega')G(\mathbf{r}',\mathbf{r};\omega'-\omega) \Big). \nonumber
\end{eqnarray}
Then we define a intermediate matrix, which does not depend on the nature of impurity potentials
\begin{eqnarray}
T(\mathbf{r},\mathbf{r}';\omega)&=&-i   \int_{-\infty}^{+\infty} \frac{d\omega'}{2\pi}   G(\mathbf{r},\mathbf{r}';\omega')G(\mathbf{r}',\mathbf{r};\omega'-\omega) \nonumber
\end{eqnarray}
After the integration, we find that $T$ reads as
\begin{eqnarray}
T(\mathbf{r},\mathbf{r}';\omega)&=&\frac{k_F}{v_F}\frac{1}{16\pi^2 r^2}e^{i|\omega| / v_Fr}\times
\begin{pmatrix}
Q &0\\
0& \sigma_z Q \sigma_z
\end{pmatrix}
\end{eqnarray}
where $\sigma_y$ is the second pauli matrix and $Q$ is defined by
\begin{eqnarray}
Q&=&\cos X\begin{pmatrix}
\frac{1}{k_Fr}+\frac{k_Fr}{k^2_Fl^2}    & \frac{k_F p_-}{k^2_F l^2}  \\
-\frac{k_F p_+}{k^2_F l^2}   &\frac{1}{k_Fr}-\frac{k_Fr}{k^2_Fl^2}  
\end{pmatrix} \nonumber\\
&+&\sin X\begin{pmatrix}
\frac{k^2_Fr^2}{2k^4_Fl^4}   &  \frac{k_F p_-}{k^2_F l^2}\frac{1}{k_Fr} \\
-\frac{k_F p_+}{k^2_F l^2}\frac{1}{k_Fr}  & \frac{k^2_Fr^2}{2k^4_Fl^4} 
\end{pmatrix}.
\end{eqnarray}
Here $X=2k_Fr-p_0^3 r^3/12$.
Then the polarization operator is given by
\begin{eqnarray}
\Pi(\mathbf{r},\mathbf{r}';\omega)&=&\text{tr} \Big(\hat{u} T(\mathbf{r},\mathbf{r}';\omega)\Big)
\end{eqnarray}
Since $T$ is quasi-diagonal, the inter-valley scattering components of the impurity vanish after the trace. So we only need to take $u$, $u_{xz}$, $u_{yz}$ and $u_{zz}$ into consideration, involving the matrices
\begin{eqnarray}
\Sigma_x \Lambda_z=\tau_0\otimes \sigma_x, \quad \Sigma_y \Lambda_z=\tau_0\otimes \sigma_y,\quad \Sigma_z \Lambda_z=\tau_z\otimes \sigma_z \nonumber
\end{eqnarray}
Among these impurity potentials, we find that only $u$-component gives non-zero contribution to PO in the leading order
\begin{eqnarray}
\label{result}
\Pi(\mathbf{r},\mathbf{r}';\omega) =&&	\frac{k_F}{v_F}\frac{u}{4\pi^2 r^2}e^{i|\omega| / v_Fr}\times  \\ \Bigg(
\frac{1}{k_Fr} \cos(2k_Fr-\frac{p_0^3r^3}{12}) 
&&+ \frac{r^2}{2k_F^2 l^4}\sin(2k_Fr-\frac{p_0^3r^3}{12})
\Bigg) \nonumber
\end{eqnarray}
As we can see, only the diagonal disorder $u\hat{I}$ gives the persistent Friedel oscillations. Other impurity potentials do not contribute to the polarization operator in the leading order in impurity scattering. In this way, we see that Eq.~(2) of main text will remain unchanged in the presence of generic non-magnetic impurity potentials.

\section{Derivation of effective momentums} \label{effective}
In this section, we give a derivation of effective momentums $k_{s,s'}$, mentioned by the qualitative argument from maintext, in the operator formalism (summations over Landau levels). Since $ \hat G_{K}$ and $ \hat G_{K'}$, propagators at valley $K$ and $K'$, can be connnected by $\hat\sigma_z \hat G_{K} \hat\sigma_z=\hat G_{K'}$, we will only focus on the calculation of one valley, $K$-valley.  

Under a magnetic field, the free Dirac electrons are transformed into ladders of Landau levels. The positive part of landau levels are given by $\omega_n=\sqrt{2n} v_F/l$. Here $v_F$ is the Fermi velocity and $l$ is the magnetic length. The corresponding wavefunction at $K$-valley is given by $\psi_{n,k_x}(\mathbf{x})=\frac{1}{\sqrt{2}}\Big(\varphi_{n-1,k_x}(\mathbf{x}),- \varphi_{n,k_x}(\mathbf{x})\Big)$, where $\varphi_{n,k_x}(\mathbf{x})$ is the wavefunction of $n^{\text{th}}$ Landau level of 2D electrons.
Considering the positive Fermi energy $E_F$, we only need to consider the Landau levels around $E_F$.
Then from definition of propagators, we get the following expression
\begin{eqnarray}
G_K^{s,s'}(\mathbf{x},\mathbf{x'};\omega)\simeq\int \frac{dk_x}{2\pi} \sum_{n} \psi^s_{n,k_x} (\mathbf{x}) \psi^{s',*}_{n,s,k_x}(\mathbf{x'})\nonumber\\
\times \frac{1}{\omega- \omega_n+i\delta \Theta(\omega-E_F)}.
\end{eqnarray}
Here $s,s'=\pm $ refer to $A/B$ sublattice and $\Theta(x)$is the step function. The off-diagonal propagtors could be expressed in terms of diagonal ones via the following expression
\begin{eqnarray}
G^{s,s'}_K(\mathbf{x},\mathbf{x'},\omega)=\frac{ l^2  \omega  {p}_{s'}}{ v_F  r^2 } \left[ G^{11}_K(\mathbf{x},\mathbf{x'},\omega)-G^{22}_K(\mathbf{x},\mathbf{x'},\omega)\right] \nonumber \\ \label{16}
\end{eqnarray}
with $\tilde{p}_\pm= {\pm(y-y')-i(x-x')} $. The derivation (see Ref.~\onlinecite{prb18Rusin} )  mainly involves the properties of Lagaurre polynomials. Thus we could only focus on the Landau summations for diagonal propagators. One puts the expression of $\psi_{n,k_x}(\mathbf{x})$ and finds
\begin{eqnarray}
G_K^{s,s }(\mathbf{x},\mathbf{x'};\omega)\simeq \frac{1}{2}\int \frac{dk_x}{2\pi} \sum_{n} \varphi_{n- (s+1)/2,k_x} (\mathbf{x})\nonumber \\ \varphi^{ *}_{n- (s+1)/2,s,k_x}(\mathbf{x'}) \frac{1}{\omega- \omega_n}.  
\end{eqnarray}
One can transfer the index $s$ into  $(\omega- \omega_n)^{-1}$ by variable change $n\rightarrow n- (s+1)/2$. This method yields 
\begin{eqnarray}
G_K^{s,s }(\mathbf{x},\mathbf{x'};\omega)\simeq \frac{1}{2}\int \frac{dk_x}{2\pi} \sum_{n} \varphi_{n,k_x} (\mathbf{x}) \varphi^{ *}_{n,s,k_x}(\mathbf{x'}) \nonumber \\
 \frac{1}{\omega- \omega_{ n+ (s+1)/2}}.
\end{eqnarray}
We expand $\omega_{ n+ (s+1)/2}=\sqrt{2n+(s+1)}v_F/l$ around the Fermi energy and find
\begin{eqnarray}
\omega_{ n+ (s+1)/2}  \simeq E_F \Big(1+\frac{s}{2k_F^2 l^2}\Big)+(\delta n+1/2 )\frac{ E_F }{  k_F^2l^2}   \nonumber
\end{eqnarray}
Here it introduce an effective Fermi energy $E_F (1+s(2k_F^2 l^2)^{-1})$ and effective cyclotron frequency $\omega_0=E_F/ k_F^2l^2$.  Subsequently, the effective Fermi energy introduces an effective momentum by $k_{s,s}=E^s_F/v_F$. One immediately see $k_{s,s}=k_F (1+s(2k_F^2 l^2)^{-1})$.

One could use Eq.~(\ref{16}) to find that effective momentum in the off-diagonal Green functions is exactly the Fermi momentum. It indicates that $k_F^{s,s'}=k_F$, if $s\neq s'$. Combining two facts, one could conclude that  
\begin{eqnarray}
k_{s,s'} =k_F \Big(1+\frac{s+s'}{4k_F^2 l^2}\Big)
\end{eqnarray}

\section{Details of  the calculation of the matrix elements $\vert M_{s,s,k}^{n,n'}\Big\vert^2$} \label{matrix}
We start from the
integral representation of the  Laguerre polynomials
\begin{equation}
\!L_m^n(x)\!=\!\frac{1}{2\pi} \int\limits_0^{2\pi} \frac{d\theta}{(1-e^{i\theta})^{n+1}} \exp \left\{\frac{xe^{i\theta}}{e^{i\theta}-1}-im\theta\right\}.
\end{equation}
In the vicinity of the  Kohn anomaly, $k\approx 2k_F$, we have $x\gg 1$.
Under this condition, the major contribution to the integral
Eq. (\ref{representation}) comes from the vicinity of $\theta =\pi$.
Substituting $\theta=\pi+\psi$ into the integrand and expanding with respect
to $\psi$, we get
\begin{equation}
\label{expansion}
L_m^n(x)\approx \int \frac{d\psi}{2\pi} \frac{1}{2^{n+1}}\exp\Big[\frac{x}{2}+i\pi m + i\phi_m^n(\psi)     \Big],
\end{equation}
where the phase $\phi_m^n(\psi)$ is given by
$\phi_m^n(\psi)=\left( \frac{x}{4}-m-\frac{n+1}{2}  \right)\psi +\frac{x\psi^3}{48}$.
The last term in the brackets originates from the denominator in Eq. (\ref{representation})
expanded and exponentiated. This term is important since $n\approx N_F$ and, thus, $n\gg 1$.

From Eq. (\ref{expansion}), and the form of $\phi_m^n(\psi)$, we conclude that the integrand in the
product $L^{n-n'}_{n'-1} (x) L^{n'-n}_{n-1}(x)$ contains the phase
\begin{eqnarray}
\label{phi1}
 \phi_{n'-1}^{n-n'}(\psi)+\phi_{n-1}^{n'-n}(\psi')=&&\Bigl(\frac{x}{4}-\frac{n+n'-1}{2}   \Bigr)(\psi+\psi')\nonumber\\&&+\frac{x}{48}\left(\psi^3+\psi'^3   \right), \label{6}
\end{eqnarray}
while the integrand in the product  $L^{n-n'}_{n'-1} (x) L^{n'-n}_{n}(x)$ contains the phase
\begin{eqnarray}
\label{phi2}
&&\phi_{n'-1}^{n-n'}(\psi)+\phi_{n}^{n'-n}(\psi')=\Bigl(\frac{x}{4}-\frac{n+n'}{2}   \Bigr)(\psi+\psi')\nonumber\\
&+&\frac{1}{2}\left(\psi-\psi'\right)+\frac{x}{48}\left(\psi^3+\psi'^3   \right). \label{7}
\end{eqnarray}
With the help of Eqs. (\ref{6}), (\ref{7}) one gets the integral representation for
the square of the matrix element,
which is the Eq.~(\ref{simp_M}) in the main text.
\section{Details of obtaining the integral representation of the polarization operator} \label{integral}

Since the main contribution to the sum in Eq.~(3) comes from $n$ and $n'$ close to $N_F$,
it is convenient to introduce the new variables
$m=N_F-n$, $m'=-N_F+n'$,
so that the relevant $m$, $m'$ are much smaller than $N_F$.
Note also, that the first terms in Eqs. (\ref{6}), (\ref{7})
measure the proximity of the momentum, $k$, to the Kohn anomaly $k\approx 2k_F$.
Indeed, with $x=\frac{k^2l^2}{2}$, we have
$\frac{x}{4}- \frac{n+n'}{2}\approx \frac{k_F l^2}{2}\delta k-\frac{m'-m}{2}$,
where $\delta k=k-2k_F$. On other hand, the cubic terms in Eqs. (\ref{6}), (\ref{7}) are the smooth
functions of $k$, so it is sufficient to replace $x$ by $\frac{k_F^2l^2}{2}$ in these terms.
Using these notations in Eqs. (\ref{simp_M}), we rewrite the static polarization operator as a sum over $m$ and $m'$
 \begin{widetext}
\begin{eqnarray}
&&\label{Pi_1}
\Pi(k)=\sum_{m,m'}
\frac{n_F(\omega_{N_F+m'})-n_F(\omega_{N_F-m})}{\omega_{N_F+m'}-\omega_{N_F-m} }
\times  \frac{1}{4\pi^3 l^2}\int \frac{d\psi d\psi'}{4}
\exp   \Big[ i\frac{k^2_Fl^2}{24}\left(\psi^3+\psi'^3\right) \Big]\times\\
& & \Biggl[ \sum_{\nu=\pm 1}\exp    \left\{i\Big( \frac{k_F l^2 \delta k}{2}-\frac{m'-m+\nu}{2} \Big)(\psi+\psi')\right\}  -2\exp    \left\{i\Big( \frac{k_F l^2 \delta k}{2}-\frac{m'-m}{2}  \Big)(\psi+\psi') +i\frac{\psi-\psi'}{2}\right\} \Biggr]. \nonumber
\end{eqnarray}
\end{widetext}
Subsequent steps rely on the relative smallness of $m$, $m'$.
Using this smallness, we expand  $\omega_{N_F-m}-\omega_{N_F+m'}$ as
$\omega_{N_F-m}-\omega_{N_F+m'}\approx -\frac{\sqrt{2}v_F}{2l}(m+m')$,
and replace the difference $n_F(\omega_{N_F+m'})-n_F(\omega_{N_F-m})$ by
$n_F(\omega_{N_F+m'})-n_F(\omega_{N_F-m})\approx n_F\Big(\!\epsilon_F+\frac{\sqrt{2}v_F}{2l}m'\!\Big)\!-\!n_F\left(\!\epsilon_F-\frac{\sqrt{2}v_F}{2l}m\!\right)$.
Finally, we extend the summation over $m$ and $m'$ to $\pm\infty$.

The above simplifications allow to perform the summation over $m$ and $m'$ explicitly.
This is achieved with the help of the identity in Eq.(\ref{identity}).
Upon applying this identity, the polarization operator Eq. (\ref{Pi_1}) acquires the form
\begin{eqnarray}
\label{Pi_2}
\Pi(k)
&=&-\frac{ k_F  }{ 2\pi^2 v_F^2} \int \frac{d\psi d\psi'}{4\vert\psi+\psi'\vert}
\exp \Big[i\frac{k^2_Fl^2}{24}(\psi^3+\psi'^3) \Big]\nonumber\\&& \times \Biggl[ \sum_{\nu=\pm 1}\exp    \left\{i\Bigl( \frac{k_F l^2 \delta k}{2}-\frac{\nu}{2} \Bigr)(\psi+\psi')\right\} \nonumber\\
& -&2\exp    \left\{i  \frac{k_F l^2 \delta k}{2}  (\psi+\psi') +i\frac{\psi-\psi'}{2}\right\} \Biggr]. 	
\end{eqnarray}
Define $2y=\psi+\psi'$,  $2z=\psi-\psi'$.
Then the double integral assumes the form
\begin{eqnarray}
\label{Pi_3}
&& \Pi(k)
=-\frac{ k_F  }{ 2\pi^2 v_F^2}   \int \frac{dy dz}{4|y|}
\exp   \left[  i\frac{k^2_Fl^2 }{12}\left(y^3+3yz^2\right) \right] \nonumber\\
&& \Biggl\{ \sum_{\nu=\pm 1}\exp    \Bigl(ik_F l^2 \delta k y - i\nu y\Bigr)   -2\exp    \Bigl(i   k_F l^2 \delta k  y + iz\Bigr) \Biggr\}.\nonumber
\end{eqnarray}
We see that the exponents in both terms are the quadratic forms of $z$.
Thus, with respect to $z$, the integrals of both terms are gaussian.
The result of the integration reads
\begin{eqnarray}
&&\Pi( k)
\approx-\frac{1  }{ 4 \pi^{3/2}v_F^2 l } \int\limits_a^{\infty}\frac{dy}{y^{3/2}}   \Biggl\{ \sum_{\nu=\pm 1}\cos \Biggl[  \left( k_F\delta k l^2-\nu\right)  y  \nonumber\\&& +\frac{k^2_Fl^2y^3}{12}+\frac{\pi}{4}\Biggr]  
- 2\cos \Biggl[    k_F \delta k l^2 y+\frac{k^2_Fl^2y^3}{12} -\frac{1}{k_F^2l^2y} +\frac{\pi}{4}  \Biggr] \Biggr\}. \nonumber
\end{eqnarray}
Then one could express it as the sum of two terms $\Pi_1(k)$ and $\Pi_2(k)$ to reproduce Eq.~(\ref{Pi_one}) and Eq.~(\ref{Pi_second}) in the maintext.  

\section{Relation bewteen the Hartree potential and the static polarization opeartor} \label{Hartree}
The modulation
of the electron density, $\delta n(\mathbf{r})$, around the impurity determines the Hartree potential by
\begin{eqnarray}
V_H(r)=\int d^2r_1 V( \mathbf{r}-\mathbf{r}_1 ) \delta n(\mathbf{r}_1)
\end{eqnarray}
Here $V( \mathbf{r}-\mathbf{r}_1 )$ is the screened Coulomb potential. Since the modulation originates from the backscattering of fermions,   $\delta n(\mathbf{r}_1)$ oscillates with the momentum $2k_F$. This implies that the Fourier component $\delta n(\mathbf{k})$ is nonzero only when $|\mathbf{k}|=2k_F$. Upon the Fourier expansion of $V( \mathbf{r}-\mathbf{r}_1 )=\int d^2k (2\pi)^{-2} e^{i\mathbf{k}( \mathbf{r}-\mathbf{r}_1 )} V(\mathbf{k})$, one obtains 
\begin{eqnarray}
V_H(r)=\int \frac{d^2k}{(2\pi)^2}     e^{i\mathbf{k}  \mathbf{r}   } V(\mathbf{k}) \delta n(\mathbf{k})
\end{eqnarray}
Here we use the expression of $\delta n(\mathbf{k})$ that $\int d^2r_1 e^{-i\mathbf{k}  \mathbf{r}_1 }  \delta n(\mathbf{r}_1)=\delta n(\mathbf{k})$. Since the Coulomb potential is rotational-invariant, $V(\mathbf{k})=V({k})$. Thus the expression above is further simplified to be
\begin{eqnarray}
V_H(r)=V(2k_F) \int \frac{d^2k}{(2\pi)^2}     e^{i\mathbf{k}  \mathbf{r}   }  \delta n(\mathbf{k})
\end{eqnarray}
One use the Fourier expansion of $\delta n(\mathbf{r})$ and immediately see
\begin{eqnarray}
V_H(\mathbf{r})=V(2k_F)  \delta n(\mathbf{r})
\end{eqnarray}
In the Born approximation, the modulation of density $\delta n(r)$ is estimated by
$\delta n(\mathbf{r})=g \Pi(r)$.
Here $g=u\int d^2r V_{imp}(r)$ and $\Pi(r)$ is the static polarization operator in real space. Finally, one obtains the relation between the static polarization operator and the Hartree potential
\begin{eqnarray}
V_H(\mathbf{r})=g V(2k_F)     \Pi(r)
\end{eqnarray}
\section{Details of the calculation of $\Pi'_1(k)$} \label{momentum}
The first step is to redefine the variable inside the integral of Eq.~(15) of the main text
$$t=(k_F l)^{2/3} y.$$ Then Eq.~(15) of the main text is now written as the integral over $t$
\begin{eqnarray}
 \Pi_{1}(k)=&&\frac{(k_Fl)^{1/3}  }{ 4 \pi^{3/2}v_F l } \int\limits_{a'}^{\infty}\frac{dt}{t^{3/2}} \Biggl\{ \sum_{\nu=\pm 1}\cos \Biggl[  \left(  \frac{\delta k}{p_0}+\nu\sqrt{\frac{p_0}{k_F}}\right) t\nonumber\\&&+\frac{t^3}{12} +\frac{\pi}{4}\Biggr] - 2\cos \Biggl[    \frac{\delta k}{p_0} t+\frac{t^3}{12} +\frac{\pi}{4}  \Biggr] \Biggr\}.  
\end{eqnarray}
with the new cut-off $a'=(k_F l)^{2/3} a$. The next step is to take the first derivatve of $\Pi_1(\delta k,0)$ versus $\delta k$
\begin{eqnarray}
 \Pi'_{1}(k)=&&-\frac{(k_Fl)^{1/3}  }{ 4 \pi^{3/2}p_0 lv_F } \int\limits_{a'}^{\infty}\frac{dt}{\sqrt{t}}
\Biggl\{ \sum_{\nu=\pm 1}\sin \Biggl[  \left(  \frac{\delta k}{p_0}+\nu\sqrt{\frac{p_0}{k_F}}\right) t\nonumber\\&&+\frac{t^3}{12} +\frac{\pi}{4}\Biggr]- 2\sin \Biggl[    \frac{\delta k}{p_0} t+\frac{t^3}{12} +\frac{\pi}{4}  \Biggr] \Biggr\}. 
\end{eqnarray}
Here we use the integral expression of product of Airy functions 
\begin{eqnarray} 
\text{Ai}(z)\text{Bi}(z)=\frac{1}{2\pi^{3/2}}\int_0^{\infty}dt  \frac{1}{\sqrt{t}} \sin\Big(zt+{\pi}/{4}+{t^3}/{12}\Big). \nonumber
\end{eqnarray}
Then we write $\Pi'_{1}(\delta k,0)$ as summation of three Airy functions
\begin{eqnarray}
\label{prime_pi1}
&&\Pi'_{1}(k)=-\frac{(k_Fl)^{1/3}   }{ 2p_0 l v_F }  \Bigl\{\sum_{\nu=\pm}\text{Ai}\Big(  \frac{\delta k}{p_0}  +\nu\sqrt{\frac{p_0}{k_F}} \Big)\nonumber\\
&&\times\text{Bi}\Big(  \frac{\delta k}{p_0} +\nu\sqrt{p_0/k_F}\Big)   -2\text{Ai}\Big(  \frac{\delta k}{p_0} \Big)\text{Bi}\Big( \frac{\delta k}{p_0} \Big)
\Bigr\}. 
\end{eqnarray}
Since $\sqrt{p_0/k_F}\ll 1$ (weak magnetic field), one can take Taylor Expansion of the product of Airy functions $$\text{Ai}(x+\nu\sqrt{p_0/k_F} )\text{Bi}(  x+\nu\sqrt{p_0/k_F}) $$ around $x=\delta k/p_0 $. Define $F(x)\equiv \text{Ai}( x )\text{Bi}(  x) $. Up to the second order perturbation,
Eq.~(\ref{prime_pi1}) is rewritten as
\begin{eqnarray}
&&\Pi'_{1}(k)\approx -\frac{(k_Fl)^{1/3}   }{ 2p_0v_F }\times\Bigl\{\sum_{\nu=\pm}\Big[ F(x) +\nu\sqrt{p_0/k_F}  F'(x) \nonumber\\
&& +\frac{p_0/k_F}{2}F''(x)\Big]-2F(x)
\Bigr\}\Big|_{x=\delta k/p_0}.
\end{eqnarray}
Here, the leading term cancels since $$2\text{Ai}(  \delta k /p_0 )\text{Bi}(  \delta k /p_0)-2\text{Ai}(  \delta k /p_0 )\text{Bi}(  \delta k /p_0)=0,$$ while the first order perturbation $\propto  \sqrt{p_0/k_F}$ also vanishes due to the fact
$$\sum_{\nu=\pm}\nu\sqrt{p_0/k_F} \Bigl( \text{Ai}(  \delta k /p_0 )\text{Bi}(  \delta k /p_0) \Bigr)'=0. $$ So the left term is the second order perturbation in the Taylor Expansion
\begin{eqnarray}
\Pi'_{1}(k)&=&-\frac{(k_Fl)^{-2/3}   }{2 v_F }     \left(\text{Ai} (  \delta k /p_0 )\text{Bi} (  \delta k /p_0 ) \right)''
\end{eqnarray}
Upon using $\sqrt{p_0/k_F}=(k_Fl)^{-2/3}$, we reach
\begin{eqnarray}
\Pi'_{1}(k)&=&-\frac{\sqrt{p_0/k_F}  }{2 v_F }     \left(\text{Ai} (  \delta k /p_0 )\text{Bi} (  \delta k /p_0 ) \right)'' \label{final_pi_one}
\end{eqnarray}

 	\bibliography{document}

\end{document}